# Optical Signals Synthesized from an Optical Lattice Clock with Uncertainty of 1.3×10⁻¹⁷

Yuan Yao, Jinpeng Zou, Haosen Shi, Hongfu Yu, Longsheng Ma, and Yanyi Jiang

State Key Laboratory of Precision Spectroscopy, East China Normal University, Shanghai 200062, China

yyao@lps.ecnu.edu.cn

**Abstract.** We report an accurate optical frequency synthesizer, which can generate single-frequency laser light with high frequency stability and accuracy at desired frequencies over a wide optical region. The frequency of the output signal is divided from an $^{171}$Yb optical lattice clock via an accurate optical frequency divider based on an optical frequency comb. Therefore, the output of the optical frequency synthesizer inherits the frequency accuracy from the Yb optical clock. The frequency uncertainty of the $^{171}$Yb optical lattice clock is evaluated to be 6.8 mHz, corresponding to a fractional frequency uncertainty of $1.3\times10^{-17}$, mainly limited by the blackbody radiation shift and the lattice-induced light shift

## 1 Introduction

Frequency synthesizers in the optical region, or called optical frequency synthesizers (OFSs), have been dreamed since 1960s. Researchers have built frequency chains to measure the frequencies of optical frequency standards relative to the primary frequency standards at ~9 GHz [1]. However, due to the complexity of the frequency chain, the measurement uncertainty is limited at $10^{-10}$–$10^{-12}$. Optical frequency combs invented in around 2000 [2,3] significantly reduce the additional frequency uncertainty in optical frequency transfer down to $10^{-19}$ [4]. Nowadays, the additional frequency uncertainty during frequency transfer has been pushed down to $10^{-21}$ [5,6] or even below [7]. The invention of optical frequency combs paves the way for OFSs.

An OFS can be referenced to a microwave frequency standard [8] or an optical frequency standard [9-11]. Significant progress of optical atomic clocks has been made in the past decade: the long-term frequency instability and uncertainty of optical atomic clocks have approached to $10^{-18}$ or even $10^{-19}$ [12-14], outperforming microwave atomic clocks by at least two orders of magnitude. Using OFSs referenced to optical atomic clocks as a powerful tool, researchers can realize measurement and detection with unprecedented precision and sensitivity, such as searching for possible variations of fundamental constants [15] and exploration of dark matter [16]. Most importantly, the SI unit of the second will be redefined based on optical atomic clocks [17,18]. Thereby, OFSs will be indispensable tools to convert the frequency of the new definition of the SI second to any desired frequency in the optical region or even in the microwave region.

In this paper, we report an accurate OFS by referencing to an Yb optical lattice clock. With an optical frequency divider, the frequency of the Yb optical lattice clock is accurately divided to an output laser. The frequency instability and uncertainty induced during optical frequency division is better than $5\times10^{-18}$ at 1 s averaging time and $10^{-21}$, respectively. Therefore, the frequency instability and uncertainty of the output laser is mainly limited by the Yb optical lattice clock The overall frequency uncertainty of the Yb optical lattice clock is $1.3\times10^{-17}$, mainly limited by the BBR shift and lattice light shift. Further improvement of reducing the frequency uncertainty of our Yb optical lattice clock down to $10^{-18}$ is briefly discussed.

## 2 Outline of an Optical Frequency Synthesizer

The output of the OFS can be a single-frequency wavelength-tunable laser such as a Ti:Sapphire continuous wave (c.w.) laser [11,19], or a laser at a desired wavelength. The frequency of the output laser light is accurately set by an optical frequency divider (OFD) as

$$\nu_{\rm o} = \nu_{\rm Yb} \div R_{\rm x}, \tag{1}$$

where $\nu_{\rm Yb}$ is the frequency of a Yb optical lattice clock and $R_{\rm x}$ is a preset divisor. The Yb optical lattice clock here is used as the frequency reference. The divisor is tunable, and it can be preset to an arbitrary desired number.

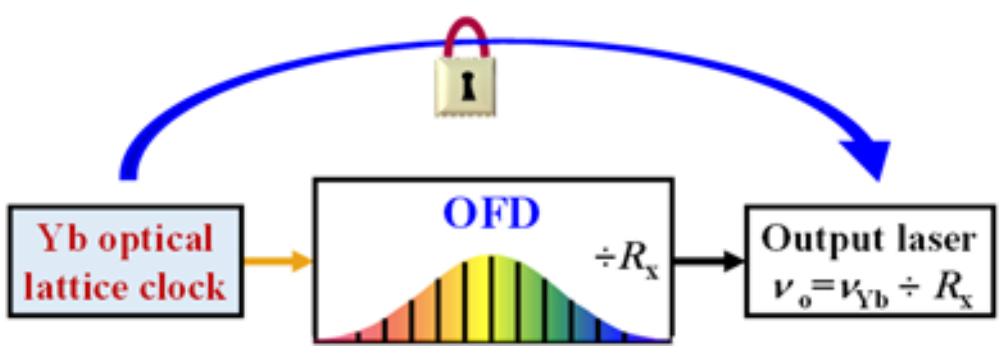

**Fig. 1.** Block diagram of an optical frequency synthesizer referenced to an Yb optical lattice clock. The frequency of the output laser is set by the optical frequency divider (OFD) as $\nu_{\rm o}=\nu_{\rm Yb}\div R_{\rm x}$.

The key element of the OFD is an optical frequency comb. Here we employ a Ti:Sapphire mode-locked femtosecond laser to generate an optical frequency comb with a repetition rate $f_{\rm r}$ of 1 GHz. The carrier-envelope offset frequency ($f_0$) is detected using a collinear $1f$-$2f$ self-referencing set-up [20]. We stabilize $f_{\rm r}$ and $f_0$ to radio frequency (RF) signals synthesized from a hydrogen maser, which has a frequency instability of $10^{-13}$ at 1 s averaging time. Certainly, we can also stabilize the comb to a cavity-stabilized laser [5] or a rubidium clock [19] as we did in previous work. In spite of the noisy comb, the output laser light is phase-locked to the Yb optical lattice clock with the technique of optical frequency divider [5].

The working principle of the OFD can be found in Ref. [5] in details. Here we summarize its main idea. First, the beat notes between the reference laser light (the Yb optical lattice clock), the output laser light against their nearby comb lines are detected as

$$f_{\rm b1} = \nu_{\rm Yb} - \nu_{\rm N1} = \nu_{\rm Yb} - (N_1 f_{\rm r} \pm f_0),$$

$$f_{\rm b1}^{*} = \nu_{\rm N1+1} - \nu_{\rm Yb} = (N_1 + 1)f_{\rm r} \pm f_0 - \nu_{\rm Yb},$$

$$f_{\rm b2} = \nu_{\rm out} - \nu_{\rm N2} = \nu_{\rm out} - (N_2 f_{\rm r} \pm f_0), \quad (2)$$

where $N_1$ and $N_2$ are integers, $\nu_{\rm Yb}$ and $\nu_{\rm out}$ are the frequencies of Yb optical clock and the output laser light, respectively, and $\nu_{\rm N}$ is the frequency of the $N$th comb line. In order to understand easily, we use the plus sign before $f_0$ in the following text.

To subtract the comb frequency noise, the transfer oscillator scheme [21] is adopted. We mix the above beating signals with $f_0$ to subtract $f_0$. Meanwhile, in order to set the output frequency of the OFD to an arbitrary number, we also mix $f_{\rm b2}$ with a frequency tunable signal of $f_{\rm tune}$ before sending to direct digital synthesizers (DDSs) for subtraction of $f_{\rm r}$. As a result, a comb-noise free error signal for phase-locking the output laser light to $\nu_{\rm Yb}$ is generated as

$$\Delta = \frac{f_{\rm b1}+f_0}{M_1} - \frac{f_{\rm b2}+f_0-f_{\rm tune}}{M_2} = \frac{\nu_{\rm Yb}-N_1 f_{\rm r}}{M_1} - \frac{\nu_{\rm out}-N_2 f_{\rm r}-f_{\rm tune}}{M_2} = \frac{\nu_{\rm Yb}}{M_1} - \frac{\nu_{\rm out}-f_{\rm tune}}{M_2} \;, \quad (3)$$

where $M_1$ and $M_2$ are the divisors of the DDSs, and they satisfy $N_1/M_1 = N_2/M_2$ to subtract $f_{\rm r}$.

In order to make the output of the OFD only related to $\nu_{\rm Yb}$, we use the transfer oscillator scheme again to obtain an optically-referenced RF time base at about 10 MHz, which is

$$f_{\rm time} = \left[\frac{f_{\rm b1}+f_0}{K_1} + \frac{f_{\rm b1}^{*}-f_0}{K_2}\right]\frac{1}{K_3} = \left[\frac{\nu_{\rm Yb}-N_1 f_{\rm r}}{K_1} + \frac{(N_1+1)f_{\rm r}-\nu_{\rm Yb}}{K_2}\right]\frac{1}{K_3} = \frac{\nu_{\rm Yb}}{K_3}\left(\frac{1}{K_1} - \frac{1}{K_2}\right) = \frac{\nu_{\rm Yb}}{k}, \quad (4)$$

where $K_1$ and $K_2$ are the divisors of anther two DDSs, and they satisfy $N_1/K_1 = (N_1+1)/K_2$. Then, we use $f_{\rm time}$ to generate $f_{\rm tune} = f_{\rm time}/K_4 = \nu_{\rm Yb}/K$, where $1/K = (1/K_1 - 1/K_2)/(K_3K_4)$.

As long as the output laser light is phase-locked to $\nu_{\rm Yb}$, the output laser frequency is

$$\nu_{\rm out} = \frac{M_2}{M_1}\nu_{\rm Yb} + \frac{\nu_{\rm Yb}}{K} = \frac{\nu_{\rm Yb}}{R}, \quad (5)$$

where the divisor $1/R = M_2/M_1 + (1/K_1 - 1/K_2)/(K_3K_4)$, only determined by the divisors of the DDSs.

We have tested the instability of $R$ on the order of $10^{-18}$ at 1 s averaging time and $10^{-21}$ at $10^4$ s averaging time, and uncertainty of $R$ better than $5\times10^{-21}$ when the comb is frequency stabilized to a rubidium clock, a hydrogen maser or a cavity-stabilized laser [5,7,19], which is negligible when synthesizing optical frequencies from the state-of-the-art optical atomic clocks.

We employ a single-frequency, Ti:Sapphire c.w. laser with more than 1 W output power as the main output of the OFS [10]. The laser wavelength can be coarsely set to a desired value in the range of 700-1000 nm by tuning a birefringence filter. Then its frequency is exactly controlled by adjusting the piezo inside the laser cavity as a slow servo and the driving frequency of a double-passed acoustic-optic modulator (AOM) at the output of the laser as a fast servo. As long as the output laser is phase-locked to the Yb optical clock, we have $\nu_{\rm out} = \nu_{\rm Yb}/R_{\rm x}$, where $R_{\rm x}$ is a preset divisor.

Meanwhile, a laser light at 759 nm, which is used as the lattice laser of the Yb optical clock, beats against the same comb. This beat note together with the signals of $f_{\rm b1}$, $f_0$ are used to realize $\nu_{lat} = \nu_{Yb}/R_1$, where $\nu_{lat}$ is the frequency of the lattice laser and $R_1$ is a divisor to make the lattice light close to the magic wavelength to largely reduce the lattice light shift [22]. In addition, the OFS outputs an optically-referenced RF time base of $f_{time} = \nu_{Yb}/k$ at about 10 MHz with the same optical frequency comb， as mentioned above[5].

## 3 Ytterbium Optical Lattice Clock

Since the additional frequency noise and uncertainty induced by the optical frequency divider is nearly two orders of magnitude lower than the state-of-the-art optical atomic clocks, the frequency stability and uncertainty of the OFS are determined by those of the Yb optical lattice clock. In this section, we will describe the experimental setup .

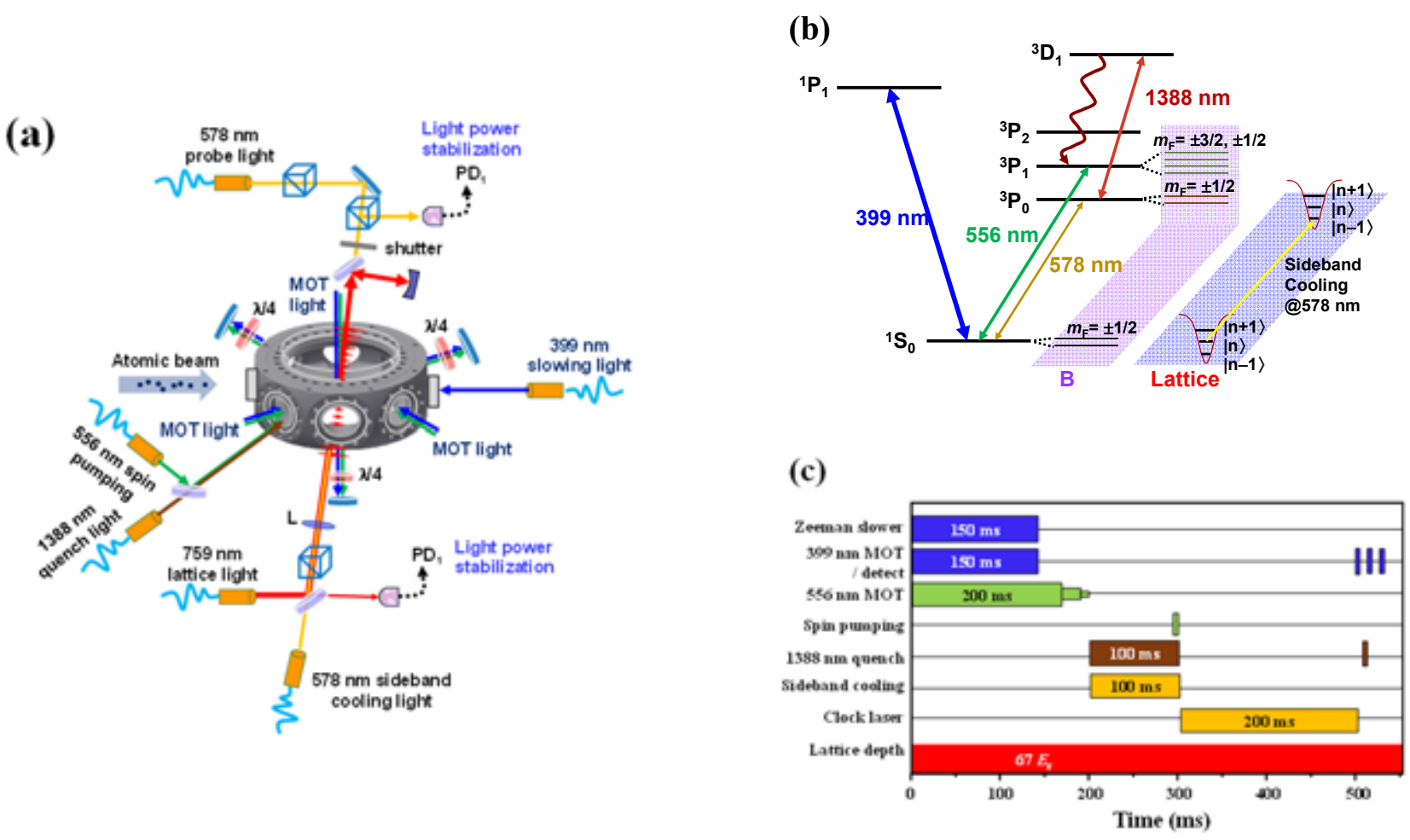

**Fig. 2.** Schematic diagram of Yb optical lattice clock. (a) Schematic layout for atomic interrogation, (b) the related energy levels, and (c) timing chart for laser cooling, state preparation and spectroscopy.

The experimental diagram of the Yb optical lattice clock is shown in Fig. 2(a). It is similar to our previous work [23]. Figure 2(b) and (c) show the related energy levels and the timing sequence of the Yb optical clock.

The atoms are loaded into a one-dimensional optical lattice at the magic wavelength of 759 nm. The operating trap depth of the lattice is ~$80E_r$, and the recoil energy is $E_r=(h\nu_{lat}/c)^2/(2m_{Yb})$, where $h$ is the Planck constant, and $m_{Yb}$ is the mass of $^{171}$Yb atom. The diameter of the optical lattice is about 110 μm.

To further cool the atoms, sideband cooling is employed [24]. In order to evaluate the atomic temperature in the longitudinal lattice dimension as well as the trap depth, we perform motional sideband spectroscopy. Figure 3(a) shows the measured motional sideband spectra, from which the axial atomic temperature can be calculated [25]. When there is no sideband cooling the atomic temperature is about 2.5 μK, while it is 0.8 μK with sideband cooling. At a lower temperature, atoms have a lower average axial motional quantum number, and thus a lower lattice light shift [24], as discussed in Section 4.

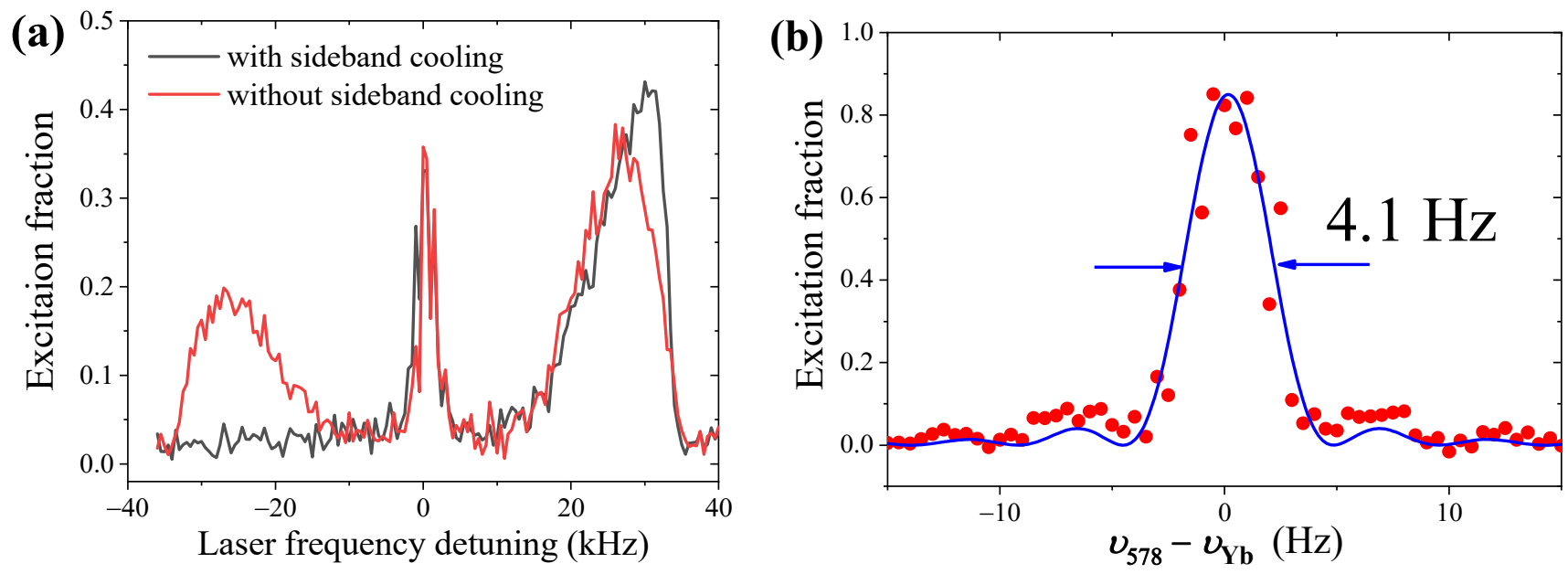

**Fig. 3.** (a) Motional sideband spectrum. (b) Rabi spectrum of Yb clock transition (no averaging) when the atoms are prepared in a single spin state. The blue solid line is the fit to the data.

For implement of the close loop of the optical lattice clock, Rabi interrogation is performed by applying a $\pi$ pulse of 578 nm probe light to excite the transition of $^1S_0$–$^3P_0$. The atoms are in a single spin state, with optically pumping. A clock laser at 578 nm is frequency-stabilized to a 30-cm-long optical cavity, yielding a laser frequency instability of $2\times10^{-16}$ at 1 s averaging time [26]. The linear frequency drift of the cavity-stabilized clock laser is feedforward compensated, which is typically below 0.05 Hz/s. With an interrogation time of 200 ms, a Rabi spectrum with a linewidth of 4.1 Hz is obtained, as shown in Fig. 3(b).

Based on the above Rabi spectrum, we stabilize the clock laser to the atomic transition by adjusting the driving frequency of an AOM placed between the 578 nm laser and its reference cavity. An interleave measurement shows that the frequency instability of the optical clock can be $5.4 \times 10^{-16}/\sqrt{\tau}$ [23], provided that the frequency shifts of the atomic transition are kept stable.

## 4 Frequency Uncertainty Evaluation

The systematic frequency shifts and relative uncertainties of the Yb lattice clock are listed in Table 1.

**Table 1.** Systematic frequency shifts and relative uncertainties

| Effect | Shift (Hz) | Uncertainty (mHz) | Fractional uncertainty ($\times10^{-17}$) |
|---|---|---|---|
| BBR shift | –1.2624 | 5.7 | 1.10 |
| Collision shift | –0.0027 | 1.7 | 0.33 |
| Lattice shift | 0.0094 | 2.9 | 0.56 |
| 2nd Zeeman shift | –0.099 | 0.8 | 0.15 |
| Background gas shift | –0.0013 | 0.1 | 0.02 |
| Probe light shift | –0.00008 | 0.04 | 0.01 |
| DC Stark shift | 0.0002 | 0.1 | 0.02 |
| Servo error | 0 | <0.5 | <0.10 |
| Total | -1.3559 | 6.7 | 1.3 |

### 4.1 BBR Stark Shift

The BBR generally comes from the MOT chamber and other ambient non-zero temperature sources around the Yb atoms. The effective temperature $T$ of the environment are used to evaluate the BBR shift.

We installed eight high precision commercial platinum resistance thermometers (PRTs, PT100-385) on the vacuum chamber, four on the top and four at the bottom. These thermometers were calibrated at Shanghai Institute of Metrology with an uncertainty less than 30 mK in the range of 20–25 °C. We recorded temperature over 26 hours in normal operation of the Yb optical clock. According to a uniform or rectangular distribution model [27], the effective temperature of the vacuum chamber is $T$ = 297.30 K, and its uncertainty is $u_{T\text{-vac}}$ = 0.192 K.

The BBR shift can be evaluated as

$$\Delta\nu_{\text{BBR}} = -\frac{1}{2}\frac{\Delta\alpha(0)}{h}\langle E^2\rangle_T[1+\eta_{\text{clock}}(T)], \tag{6}$$

where $\Delta\alpha(0) = h\times3.62612(7)\times10^{-6}$ Hz/(V/m)$^2$ is the differential static polarizability between the two clock states, $\langle E^2\rangle_T = [8.319430(15)\text{ V/cm}]^2 \times (T/300\text{ K})^4$ is the mean-squared electric field in a BBR environment with absolute temperature $T$, and $\eta_{\text{clock}}(T) \approx 0.01745(38) \times (T/300\text{ K})^2 + 0.000593(16) \times (T/300\text{ K})^4$ is a small dynamic correction to account for frequency dependence of the state polarizabilities across the BBR spectrum [29]. The BBR shift induced by the vacuum chamber is estimated to be −1.2318(32) Hz with a fractional uncertainty of $6.2 \times 10^{-18}$.

Since the vacuum chamber is not fully enclosed by the vacuum chamber, we need to take other apertures into consideration. The effective temperature seen by the atoms through these apertures can be determined using the method described in Ref. [28]. After substituting the effective radiation temperature into Eq. (6), we can obtain the BBR shift induced by these apertures, as shown in Table 2.

**Table 2.** BBR shift uncertainty for the Yb lattice clock in normal operation

| BBR environment | Shift (Hz) | Uncertainty (mHz) | Fractional uncertainty ($\times10^{-18}$) |
|---|---|---|---|
| Vacuum chamber | −1.2318 | 3.2 | 6.2 |
| Viewport for slowing light | −0.030 1 | 4.6 | 8.9 |
| Atomic shutter | −0.000 52 | 0.5 | 1 |
| Chamber windows | 0 | 0.2 | 0.4 |
| $\eta$ | 0 | 0.4 | 0.8 |
| Temperature measurement | 0 | 0.5 | 1 |
| Total | −1.262 4 | 5.7 | 11 |

### 4.2 Lattice Light Shift

The AC Stark shift of the lattice light is modeled as [24]

$$\Delta\nu_{\text{lat}} = [\alpha'\Delta\nu_{\text{lat}} - \alpha^{qm}]\left(n_i + \frac{1}{2}\right)\sqrt{\frac{\zeta V_0}{E_{\text{r}}}}$$

$$-\left[\alpha'\Delta\nu_{\text{lat}} + \frac{3}{4}\beta\left(2n_i^2 + 2n_i + 1\right) + \alpha^{qm}(r-1)\right]\frac{\zeta V_0}{E_{\text{r}}}$$

$$+\beta(2n_i + 1)r\left(\frac{\zeta V_0}{E_{\text{r}}}\right)^{\frac{3}{2}} - \beta\left(\frac{r\zeta V_0}{E_{\text{r}}}\right)^2, \tag{7}$$

where $\alpha'$ is the slope of the differential electric dipole ($E_1$) polarizability, $\Delta\nu_{\text{lat}}$ is the lattice frequency detuning from the $E_1$ magic frequency, $V_0$ is the potential depth, $\zeta$ is lattice-intensity reduction factor due to the atomic distribution in an optical lattice with nonuniform intensity, $r$ is a correction factor characterizing the intensity imbalance between two propagating lattice beams, and $n_{\text{i}}$ is axial motional quantum number.The coefficients of $\alpha^{\text{qm}}$ and $\beta$ come from the multipolar polarizability and the hyperpolarizability, respectively.

In normal operation of our apparatus, $V_0$ = 80(2)$E_{\text{r}}$ and $n_{\text{i}}$ = 0.03(1) obtained from the motional sideband spectrum in Fig. 3(a) [25]. The reduction factor of $\zeta$ is evaluated to be 0.85(4) based on the radial temperature of $T_{\text{r}}$=1.2(2) μK, which is measured from the Doppler broadened spectrum of the clock transition in the transverse direction. The intensity of the lattice return beam is attenuated to 85(3)%, and thus $r$ is calculated to be 1.0016(7). According to Ref. [24], $\alpha'$= 25.74(54) μHz/MHz, $\beta$= −1.194(89) μHz, $\alpha^{\text{qm}}$= −1027(378) μHz, and the magic wavelength is 394 798 261.06 (1.37) MHz. Here we stabilize the lattice laser at 394 798 261 MHz using the OFD. We evaluate the scalar/tensor light shift (the part relevant to $\alpha'$) to be 0(2.2) mHz, and those due to hyperpolarizability and multipolar polarizability are 4.9(0.37) mHz and 4.5(1.7) mHz, respectively. Therefore, the overall AC Stark shift induced by the lattice light is 9.4(2.9) mHz.

### 4.3 Density Shift

The density shift is evaluated by interleaving measurements at different atomic densities. The atomic number is set by changing the duration of the slower light at 399 nm during the first-stage MOT. Using this method, it does not impact the trapping conditions, and thus the density shift is proportional to the number of atoms [31].

We measured the frequency shift to be –0.032(21) Hz when the atomic number is switched between 13$N_{\text{atom}}$ and $N_{\text{atom}}$ in each cycle, where the atomic number is $N_{\text{atom}}$≈1,000 in normal operation. In normal operation, the typical density shift is estimated to be −2.7(1.7) mHz.

### 4.4 Second-order Zeeman Shift

We apply a magnetic field to split the two π-transitions by ~500 Hz. The frequency difference between the two π-transitions is measured with a typical uncertainty of 1 Hz. It is then used to estimate a quadratic Zeeman shift to be −99.0(0.8) mHz, using the coefficient $a_{\text{Z}}$ = −1.531(2) μHz/Hz$^2$ [13].

### 4.5 DC Stark Shift

The vacuum chamber where Yb atoms are trapped is made of aluminum, it can be served as a Faraday cage against external DC electric fields. However, electric charges may accumulate on the windows, causing a static Stark shift.

Following the ideas in Ref. [31], we placed two ring electrodes in the vertical direction, and we applied a voltage $V_{DC}$ on one electrode while another electrode was grounded. We measured the DC Stark shift induced by the background electric field in the vertical direction, which is 0.2(0.1) mHz. We also measured the DC Stark shift in the horizontal direction. Although we applied 6000 V on the electrodes, we failed to resolve any observable frequency shift sensed by the atoms. After finite elements analysis, we found that the electric field in the horizontal sensed by the atoms was 850 times smaller than that in the vertical direction, assuming there was a same charge per area accumulated on the horizontal windows. Therefore, in horizontal direction the DC Stark shift induced by the background electric field is negligible.

### 4.6 Other shifts

Collision between Yb atoms and background gas induces a frequency shift of the clock transition. The background gas is dominated by $H_2$. This shift is reported to be proportional to the trap loss rate with a coefficient of $-1.64(12) \times 10^{-17}$ s [13]. Therefore, the frequency shift from background gas collision is estimated to be $-1.3(1)$ mHz using the measured trap lifetime of 6.6(1) s.

In normal operation, we keep the probe duration $t_\pi$ = 200 ms. Since the AC Stark shift induced by the probe light is measured to be $2(1)\times10^{-20}$ with $t_\pi$ = 560 ms in Ref. [13], in this work the Stark shift is estimated to be $1.6(8) \times 10^{-19}$ as the laser intensity adjusted for the pulse scales with $(1/t_\pi)^2$.

The servo error uncertainty is estimated from the average of the error signal during the campaign. It is zero with an uncertainty of $1 \times 10^{-18}$.

## 5 Conclusion

In conclusion, we report an optical frequency synthesizer with a frequency uncertainty of $1.3 \times10^{-17}$ by referencing to an Yb optical lattice clock. Using an optical frequency divider, the frequency of the Yb optical lattice clock is accurately divided to single-frequency output lasers. The frequency uncertainty of the current Yb optical lattice clock is limited by the blackbody radiation shift and the lattice light shift. In the next step toward an optical frequency synthesizer with a frequency uncertainty at the $10^{-18}$ level, we will install an electrically-grounded blackbody radiation shield inside the vacuum chamber to make temperature homogenous and to block DC electric field as well [28]. To reduce the lattice light shift, we will measure the coefficients of $\alpha'$, $\alpha^{qm}$ and $\beta$ with less uncertainty, and meanwhile we will trap the atoms in even shallower lattice sites [24].